\documentclass[aps,pre,reprint,amsmath,amssymb,longbibliography,lengthcheck,showpacs,superscriptaddress,floatfix]{revtex4-2}
\usepackage{enumerate}
\usepackage[dvipdfmx]{graphicx}
\usepackage{color}
\usepackage{siunitx}
\usepackage[maxfloats=256]{morefloats}
\begin{document}
%
%%%%%%%%%%%%%%%%%%%%%%%%%%%%%%%%%%%%%%%%%%%%%%%%%%%%%%%%%%%%%%%%%%%%%%%%%%%%%%%%%%%%%%%%%%%%%%%%%
\title{Boson peak in the dynamical structure factor of network- and packing-type glasses}
%
%%%%%%%%%%%%%%%%%%%%%%%%%%%%%%%%%%%%%%%%%%%%%%%%%%%%%%%%%%%%%%%%%%%%%%%%%%%%%%%%%%%%%%%%%%%%%%%%%
\author{Hideyuki Mizuno}
\email{hideyuki.mizuno@phys.c.u-tokyo.ac.jp}
\affiliation{Graduate School of Arts and Sciences, The University of Tokyo, Tokyo 153-8902, Japan}
\author{Emi Minamitani}
\affiliation{SANKEN, University of Osaka, Osaka 567-0047, Japan}
%
%%%%%%%%%%%%%%%%%%%%%%%%%%%%%%%%%%%%%%%%%%%%%%%%%%%%%%%%%%%%%%%%%%%%%%%%%%%%%%%%%%%%%%%%%%%%%%%%%
\date{\today}
%
%%%%%%%%%%%%%%%%%%%%%%%%%%%%%%%%%%%%%%%%%%%%%%%%%%%%%%%%%%%%%%%%%%%%%%%%%%%%%%%%%%%%%%%%%%%%%%%%%
\begin{abstract}
Glasses are structurally disordered solids that host, in addition to crystalline-like phonons, vibrational excitations with no direct phononic counterpart.
A long-standing universal signature is the excess vibrational density of states~(vDOS) over the Debye prediction, known as the boson peak~(BP), which has been extensively reported via inelastic neutron and X-ray scattering measurements of the dynamical structure factor $S(q,\omega)$.
Here we quantify the vDOS directly from dynamical-structure-factor data and clarify the microscopic origin of the BP.
We contrast two routes to extract the vDOS from $S(q,\omega)$:
(i) using high-wavenumber $q$ data beyond the Debye wavenumber $q_D$ to access predominantly incoherent scattering and recover the vDOS in a manner analogous to velocity-autocorrelation-based approaches; and
(ii) integrating $S(q,\omega)$ over the low-$q$ regime below $q_D$, which enables a decomposition of the vDOS into contributions from distinct wavenumber sectors and thereby provides direct access to the spatial character of vibrational modes.
Focusing on the second route, we demonstrate that the BP in the vDOS emerges as the spectral consequence of a dispersionless excitation band in $S(q,\omega)$.
Our main results are obtained from molecular-dynamics simulations, and we further show that the same mechanism is captured by an effective-medium theory for random spring networks, providing a unified interpretation that connects the excess vDOS to the wavenumber-resolved structure of vibrational excitations in glasses.
\end{abstract}
%
%%%%%%%%%%%%%%%%%%%%%%%%%%%%%%%%%%%%%%%%%%%%%%%%%%%%%%%%%%%%%%%%%%%%%%%%%%%%%%%%%%%%%%%%%%%%%%%%%
\maketitle
%
%%%%%%%%%%%%%%%%%%%%%%%%%%%%%%%%%%%%%%%%%%%%%%%%%%%%%%%%%%%%%%%%%%%%%%%%%%%%%%%%%%%%%%%%%%%%%%%%%
\section{Introduction}
Glasses are structurally disordered solids that, in addition to crystalline-like phonons, support vibrational excitations with no direct phononic counterpart.
A central and remarkably universal manifestation is the excess vibrational density of states (vDOS) over the Debye prediction---the boson peak (BP)---observed across a wide range of glasses~\cite{Ramos_2022,Nakayama_2002}.
Although the BP has been documented for decades, its microscopic origin remains actively debated, in part because experiments do not always access the ``bare'' vDOS $g(\omega)$ directly (with $\omega$ the angular frequency) but instead probe response functions with nontrivial matrix elements---most prominently the dynamical structure factor $S(q,\omega)$, as discussed below.
A key goal is therefore to connect the excess vDOS to experimentally accessible observables and to clarify how it is encoded in these measurements.

Experimental access to the vDOS differs sharply between colloidal glasses and atomic/molecular glasses.
In colloids, particle coordinates can be tracked directly by confocal or video microscopy, providing an almost first-principles route to vibrational modes.
From measured displacement fields one constructs the covariance matrix, from which an effective stiffness (Hessian-like) operator can be inferred, so that the normal modes and $g(\omega)$ can be obtained from its eigenvectors and eigenvalues~\cite{Kaya_2010,Brito_2010,Henkes_2012}.
This approach parallels standard molecular-dynamics (MD) simulations, where the vibrational spectrum is obtained by diagonalizing the $T=0$ Hessian (dynamical matrix)~\cite{MizunoIkeda2022}.
Using this experimental framework, low-frequency excess modes have been investigated as functions of control parameters such as the volume fraction and the coordination number~\cite{Chen_2010,Tan_2012}.

In atomic and molecular glasses, by contrast, the vDOS is typically accessed indirectly through scattering and spectroscopic probes with nontrivial matrix elements~\cite{Buchenau_1986,Yamamuro_1996,Arai_1999,Harris_2000,Monaco_2006,Niss_2007,Ruffle_2008,Baldi_2008,Monaco_2009,Baldi_2010,Baldi_2011,Ruta_2012,Baldi_2013,Chumakov_2014,Baldi_2016}.
Inelastic neutron scattering (INS) measures the dynamical structure factor $S(q,\omega)$ and provides the most direct experimental route to the vDOS, although in practice it is commonly discussed in terms of a generalized (neutron-weighted) vDOS.
Inelastic X-ray scattering (IXS) also measures $S(q,\omega)$ and, by probing primarily longitudinal density fluctuations, constrains the physics behind the vDOS by mapping acoustic dispersions $\omega(q)$ and linewidths (dampings) $\Gamma(q)$, thereby linking the BP regime to the strong damping of acoustic excitations and Ioffe--Regel--like crossovers.
Optical probes access different correlation functions (typically at $q\simeq 0$) and thus provide coupling-weighted spectra:
for Stokes Raman scattering one often writes $I_R(\omega)=C_R(\omega)\,[n(\omega)+1]\,g(\omega)/\omega$~\cite{Shuker_1970,Surovtsev_2002,Schmid_2008}, where $n(\omega)$ is the Bose--Einstein occupation factor,
while infrared absorption is frequently expressed schematically as $\alpha(\omega)=C_{\mathrm{IR}}(\omega)\,g(\omega)$~\cite{Strom_1977,Ohsaka_1998,Taraskin_2006}.
In both cases, extracting $g(\omega)$ requires modeling the coupling functions $C_R(\omega)$ or $C_{\mathrm{IR}}(\omega)$.
THz spectroscopy offers complementary information on the same low-frequency optical response in the BP range~\cite{Kabeya_2016,Mori_2020,Kyotani_2025}.

Motivated by this experimental situation, we focus on routes that connect the BP anomaly in $g(\omega)$ directly to the dynamical structure factor $S(q,\omega)$, the central experimental observable in INS and IXS.
Broadly, there are two complementary strategies to infer $g(\omega)$ from $S(q,\omega)$.
The first exploits the high-wavenumber regime $q>q_D$, where $q_D$ is the Debye wavenumber, in which incoherent contributions dominate and $S(q,\omega)$ becomes effectively single-particle-like, enabling the reconstruction of $g(\omega)$~\cite{Price_1987,Pasquarello_1998,Fultz_2010,Squires_2012}.
This strategy closely parallels the standard estimate based on the Fourier transform of the velocity autocorrelation function, widely used in MD simulations~\cite{Grest_1981,Ikeda_2013}.
The second integrates $S(q,\omega)$ over the low-$q$ regime $q<q_D$ to obtain $g(\omega)$ while retaining a decomposition into distinct wavenumber sectors, thereby providing a wavenumber-resolved link between the measured $S(q,\omega)$ and the spatial character of vibrational excitations and their contributions to the BP~\cite{Schirmacher_2006,Wyart_2010,Marruzzo_2013,Degiuli_2014,Schirmacher_2015,Mizuno_2018}.

In the present work, we quantify the vDOS $g(\omega)$ from $S(q,\omega)$ using two complementary routes.
The first route uses high-$q$ spectra to provide an accurate and experimentally convenient estimate of $g(\omega)$.
Building on the second, wavenumber-resolved route, we show that the BP anomaly in $g(\omega)$ emerges as the spectral consequence of a nearly dispersionless excitation band in $S(q,\omega)$.
Our main results are obtained from MD simulations, and we further show that the same interpretation is captured by an effective-medium theory (EMT) for random spring networks~\cite{Wyart_2010,Degiuli_2014}.
This perspective provides an experimentally grounded bridge between BP phenomenology and the physics of non-phononic excitations in disordered solids.

%%%%%%%%%%%%%%%%%%%%%%%%%%%%%%%%%%%%%%%%%%%%%%%%%%%%%%%%%%%%%%%%%%%%%%%%%%%%%%%%%%%%%%%%%%%%%%%%%
\section{System descriptions}
In this study, we use MD simulations to analyze three representative glass formers:
silica glass as a prototypical covalent network glass, and two packing-dominated (van der Waals) glasses, namely a one-component Lennard--Jones (LJ) glass and a binary soft-sphere (SS) glass.
By contrasting network-type and packing-type systems, we aim to assess the universality of the connection between the BP and a dispersionless excitation band.

%%%%%%%%%%%%%%%%%%%%%%%%%%%%%%%%%%%%%%%%%%%%%%%%%%%%%%%%%%%%%%%%%%%%%%%%%%%%%%%%%%%%%%%%%%%%%%%%%
\subsection{Silica glass}
We consider a three-dimensional model of silica~($\text{SiO}_2$) glass composed of $N_{\mathrm{Si}}$ silicon atoms and $N_{\mathrm{O}}$ oxygen atoms, with the stoichiometric constraint $N_{\mathrm{O}}=2N_{\mathrm{Si}}$ and total atom number $N=N_{\mathrm{Si}}+N_{\mathrm{O}}=3N_{\mathrm{Si}}$.
The atomic masses are denoted by $m_{\mathrm{Si}}$ and $m_{\mathrm{O}}$, with $m_{\mathrm{Si}}/m_{\mathrm{O}}=1.755$.

Interatomic interactions are modeled using the BKS family of pair potentials, originally introduced in Ref.~\cite{Beest_1990} and subsequently refined in several works~\cite{Wolf_1992,Wolf_1999,Carre_2007,Carre_2016,Sundararaman_2018}.
In the present study, we employ the SHIK parameterization~\cite{Sundararaman_2018}, which has been benchmarked against experiments and first-principles simulations and shown to reproduce thermodynamic properties, structural observables (radial distribution functions, static structure factors, and bond-angle distributions), and elastic moduli.
Related BKS-type simulations have examined the mechanical response and vibrational properties of silica glasses~\cite{Mantisi_2012,Shcheblanov_2015,Damart_2017}, and the effects of densification on the BP have also been reported recently~\cite{Hamdaoui_2025}.

The total interaction is written as the sum of a short-range part $v_S(r)$ and a Coulomb part $v_L(r)$.
The short-range contribution reads
\begin{equation}
v_S(r) = A_{\alpha \beta} \exp \left( - B_{\alpha \beta} r \right) - \frac{C_{\alpha \beta}}{r^6} + \frac{D_{\alpha \beta}}{r^{24}},
\end{equation}
where $r$ is the interparticle distance and $\alpha,\beta\in\{\mathrm{Si},\mathrm{O}\}$.
The parameters $A_{\alpha \beta}$, $B_{\alpha \beta}$, $C_{\alpha \beta}$, and $D_{\alpha \beta}$ are listed in Table~\ref{table1}.
The Coulomb term is
\begin{equation}
v_L(r)=\frac{q_\alpha q_\beta}{r},
\end{equation}
with partial charges $q_{\mathrm{Si}}=1.7755\,e$ and $q_{\mathrm{O}}=-q_{\mathrm{Si}}/2$, ensuring overall charge neutrality (here $e$ is the elementary charge).

We apply cutoffs $r_{cS}=8.0$~\AA\ and $r_{cL}=10.0$~\AA\ to $v_S(r)$ and $v_L(r)$, respectively.
For the Coulomb interaction, this cutoff is referred to as Wolf truncation~\cite{Wolf_1992,Wolf_1999}.
To avoid discontinuities at the cutoffs, both contributions are smoothed according to
\begin{equation}
\begin{aligned}
\phi_S(r) &= \left[ v_S(r) - v_S(r_{cS}) - (r - r_{cS}) v'_S(r_{cS}) \right] G_{cS}(r), \\
\phi_L(r) &= \left[ v_L(r) - v_L(r_{cL}) \right] G_{cL}(r),
\end{aligned}
\end{equation}
with
\begin{equation}
\begin{aligned}
G_{cS}(r) &= \exp \left[ - \frac{\gamma_S^2}{(r - r_{cS})^2} \right], \\
G_{cL}(r) &= \exp \left[ - \frac{\gamma_L^2}{(r - r_{cL})^2} \right],
\end{aligned}
\end{equation}
where $\gamma_S=\gamma_L=0.2$~\AA.

%%%%%%%%%%%%%%%%%%%%%%%%%%%%%%%%%%%%%%%%%%%%%%%%%%%%%%%
\begin{table}[t]
\caption{\label{table1}
{Parameters for the potential of silica glass.}
$q_\text{Si}=1.7755\,e$, $q_\text{O}=-q_\text{Si}/2$, $r_{cS}=8.0$~\AA, $r_{cL}=10.0$~\AA, and $\gamma_S=\gamma_L=0.2$~\AA.
}
\centering
\renewcommand{\arraystretch}{1.1}
\begin{tabular}{|c|c|c|c|c|}
\hline
 $\alpha$--$\beta$ & $A_{\alpha \beta}$~(eV) & $B_{\alpha \beta}$~(\AA$^{-1}$) & $C_{\alpha \beta}$~(eV\,\AA$^{6}$) & $D_{\alpha \beta}$~(eV\,\AA$^{24}$) \\
\hline
 Si--O  & $23107.8$ & $5.098$ & $139.7$ & $66.0$ \\
\hline
 O--O   & $1120.5$  & $2.893$ & $26.1$  & $16800.0$ \\
\hline
 Si--Si & $2797.9$  & $4.407$ & $0.0$   & $3423204.0$ \\
\hline
\end{tabular}
\end{table}
%%%%%%%%%%%%%%%%%%%%%%%%%%%%%%%%%%%%%%%%%%%%%%%%%%%%%%%

We note that BKS-type models are pairwise additive, comprising a short-range two-body term together with a long-range Coulomb contribution, and therefore do not explicitly include angular many-body interactions that encode covalent directionality.
More explicit descriptions of covalent bonding incorporate angular forces, notably the Vashishta potential~\cite{Vashishta_1990}, which adds three-body interactions associated with O--Si--O and Si--O--Si angles.
Nevertheless, BKS-type models implicitly capture key aspects of covalent directionality through their parameterization and have been shown to reproduce, with quantitative accuracy, the thermodynamic properties, structural observables, and elastic moduli of silica glass~\cite{Beest_1990,Wolf_1992,Wolf_1999,Carre_2007,Carre_2016,Sundararaman_2018}.

We carried out three independent MD simulations using LAMMPS~\cite{Plimpton_1995}.
The mass density was fixed at $\rho=2.20$~g/cm$^3$, and the temperature was controlled using a Nos\'e--Hoover thermostat~\cite{Nose_1984,Hoover_1985}.
Starting from randomized Si and O positions, we equilibrated the system at $T=3500$~K for $100$~ps to obtain a homogeneous liquid.
We then quenched to $T=300$~K at a rate of $1$~K/ps and equilibrated at $T=300$~K for an additional $100$~ps.
Finally, atomic velocities were set to zero and the configuration was energy-minimized to obtain the inherent structure $\vec{r}=[\vec{r}_{1},\vec{r}_{2},\ldots,\vec{r}_{N}]$.

To cover a wide range of wavenumbers $q$ and frequencies $\omega$, we considered multiple system sizes from $N=1.5\times 10^4$ up to $2.4\times 10^5$ atoms.
All reported results are averages over the three independent inherent structures.

%%%%%%%%%%%%%%%%%%%%%%%%%%%%%%%%%%%%%%%%%%%%%%%%%%%%%%%%%%%%%%%%%%%%%%%%%%%%%%%%%%%%%%%%%%%%%%%%%
\subsection{Lennard--Jones glass}
As a simple packing-type atomic glass, we consider a three-dimensional one-component LJ system that has been studied extensively in previous works~\cite{Monaco2_2009,Mizuno_2013,Shimada_2018}.
Particles interact via
\begin{equation}\label{eq.potlj}
v(r)=4\epsilon\left[\left(\frac{\sigma}{r}\right)^{12}-\left(\frac{\sigma}{r}\right)^{6}\right],
\end{equation}
where $r$ is the interparticle distance and $\sigma$ sets the particle diameter.
The potential is truncated at $r_c=2.5\sigma$.
To suppress artifacts associated with the cutoff discontinuity~\cite{Mizuno2_2016}, we employ the shifted-force form
\begin{equation}\label{eq.potsmooth}
\phi(r)=v(r)-v(r_c)-(r-r_c)\,v'(r_c),
\end{equation}
which makes both $\phi(r)$ and its derivative vanish at $r=r_c$.
All particles have equal mass $m$, and we use reduced LJ units with length $\sigma$, mass $m$, and energy $\epsilon$.
The number density is $\hat{\rho}=N/V=1.015$, where $V$ is the system volume, and the system size ranges from $N=1.6\times10^4$ to $1.024\times10^6$.

%%%%%%%%%%%%%%%%%%%%%%%%%%%%%%%%%%%%%%%%%%%%%%%%%%%%%%%%%%%%%%%%%%%%%%%%%%%%%%%%%%%%%%%%%%%%%%%%%
\subsection{Soft-sphere glass}
We further study a three-dimensional binary SS mixture analyzed in our previous works~\cite{Mizuno2_2013,Mizuno_2014,Mizuno_2016}.
Particles of types $\alpha,\beta\in\{L,S\}$ interact through a 12-inverse-power-law potential
\begin{equation}\label{eq:sspotential}
v(r)=\epsilon\left(\frac{\sigma_{\alpha\beta}}{r}\right)^{12},
\end{equation}
where $\sigma_{\alpha\beta}=(\sigma_\alpha+\sigma_\beta)/2$.
The size ratio is $\sigma_S/\sigma_L=0.7$, and we take an equimolar composition $x_{L,S}=N_{L,S}/N=1/2$ with $N=N_L+N_S$.
As for the LJ system, we use the smoothed interaction $\phi(r)$ in Eq.~(\ref{eq.potsmooth}).
All particles have the same mass $m$, and we define the length unit as $\sigma = \left(\sum_{\alpha,\beta=L,S} x_\alpha x_\beta \sigma_{\alpha \beta}^3 \right)^{1/3}$.
The number density is again $\hat{\rho}=N/V=1.015$, and we consider system sizes from $N=1.6\times10^4$ to $1.024\times10^6$.

%%%%%%%%%%%%%%%%%%%%%%%%%%%%%%%%%%%%%%%%%%%%%%%%%%%%%%%
\begin{table*}[t]
\caption{\label{table2}
{Physical quantities including elastic moduli and Debye values.}
For silica glass, the quantities are given as the mass density $\rho$~(g/cm$^3$), elastic moduli $K$ and $G$~(GPa), sound speeds $c_L$ and $c_T$~(m/s), wavenumber $q$~(\AA$^{-1}$), Debye level $A_D$~(THz$^{-3}$), and frequency $\omega$~(THz).
Note that $\nu$ denotes the Poisson ratio.
}
\centering
\renewcommand{\arraystretch}{2.0}
\setlength{\tabcolsep}{3pt}
\begin{tabular}{|c|c|c|c|c|c|c|c|c|c|c|c|c|c|c|c|c|c|}
\hline
 Glass & $\rho$ & $K$ & $K_A$ & $K_N$ & ${\displaystyle \frac{K_N}{K_A}}~(\%)$ & $G$ & $G_A$ & $G_N$ & ${\displaystyle \frac{G_N}{G_A}}~(\%)$ & $\nu$ & $c_L$ & $c_T$ & ${\displaystyle \frac{c_L}{c_T}}$ & $q_D$ & $A_D$ & $\omega_D$ & $\omega_\text{BP}$ \\
\hline
Silica & $2.20$ & $40.9$ & $172$ & $131$ & $76.3$ & $30.5$ & $104$ & $73.6$ & $73.7$ & $0.202$ & $6090$ & $3720$ & $1.64$ & $1.58$ & $0.00257$ & $10.5$ & $1.21$ \\
\hline
LJ & $1.015$ & $61.2$ & $61.7$ & $0.530$ & $0.859$ & $13.6$ & $35.8$ & $22.2$ & $61.9$ & $0.396$ & $8.84$ & $3.67$ & $2.41$ & $3.92$ & $0.000699$ & $16.3$ & $1.05$ \\
\hline
SS & $1.015$ & $40.8$ & $40.8$ & $0.00$ & $0.00$ & $6.21$ & $14.7$ & $8.50$ & $57.8$ & $0.428$ & $6.96$ & $2.47$ & $2.81$ & $3.92$ & $0.00225$ & $11.0$ & $0.798$  \\
\hline
\end{tabular}
\end{table*}
%%%%%%%%%%%%%%%%%%%%%%%%%%%%%%%%%%%%%%%%%%%%%%%%%%%%%%%

%%%%%%%%%%%%%%%%%%%%%%%%%%%%%%%%%%%%%%%%%%%%%%%%%%%%%%%%%%%%%%%%%%%%%%%%%%%%%%%%%%%%%%%%%%%%%%%%%
\section{Numerical analyses}
%
%%%%%%%%%%%%%%%%%%%%%%%%%%%%%%%%%%%%%%%%%%%%%%%%%%%%%%%%%%%%%%%%%%%%%%%%%%%%%%%%%%%%%%%%%%%%%%%%%
\subsection{Vibrational modes}
For each inherent structure $\vec{r}=[\vec{r}_{1},\vec{r}_{2},\cdots,\vec{r}_{N}]$, we perform a standard normal-mode analysis based on the $3N\times 3N$ dynamical matrix.
Solving the corresponding eigenvalue problem yields eigenvalues $\{\lambda_k\}$ and eigenvectors $\{\vec{e}_k=[\vec{e}_{k,1},\vec{e}_{k,2},\cdots,\vec{e}_{k,N}]\}$ for $k=1,2,\cdots,3N$~\cite{MizunoIkeda2022}.
The eigenfrequencies are $\omega_k=\sqrt{\lambda_k}$, and the eigenvectors are normalized as
$\vec{e}_k\cdot\vec{e}_l=\sum_{i=1}^{N}\vec{e}_{k,i}\cdot\vec{e}_{l,i}=\delta_{k,l}$,
where $\delta_{k,l}$ is the Kronecker delta.
In the following analyses based on the mode data $\{\omega_k,\vec{e}_k\}$, we discard the three translational zero modes.

To access both moderate and very low frequencies, we combine results across multiple system sizes, following the strategy of Ref.~\cite{Mizuno_2017}.
We compute all modes for the smallest system, whereas for larger systems we obtain only low-frequency modes.
We then merge the mode datasets as a function of $\omega_k$; spectra from different sizes overlap and match smoothly, extending the accessible frequency window toward lower $\omega$.
The vDOS $g(\omega)$ in Eq.~(\ref{eq.vdos}) and the dynamical structure factors $S_\alpha(q,\omega)$ in Eq.~(\ref{eq.sqomega}) are evaluated from these combined datasets, and data from different sizes are presented together.

%%%%%%%%%%%%%%%%%%%%%%%%%%%%%%%%%%%%%%%%%%%%%%%%%%%%%%%%%%%%%%%%%%%%%%%%%%%%%%%%%%%%%%%%%%%%%%%%%
\subsection{Vibrational density of states}
Given the eigenfrequencies $\{\omega_k\}$, the vDOS is computed as
\begin{equation}\label{eq.vdos}
g(\omega)=\frac{1}{3N}\sum_{k=1}^{3N}\delta(\omega-\omega_k),
\end{equation}
where $\delta$ is the Dirac delta function.

In an isotropic elastic continuum, low-frequency excitations are phonons with linear dispersions $\omega=c_T q$ and $\omega=c_L q$, where $c_T$ and $c_L$ are the transverse and longitudinal sound speeds, respectively~\cite{Ashcroft_1976}.
Counting phonon states leads to the Debye prediction,
\begin{equation}\label{equofdebye}
g_D(\omega)=A_D\omega^{2}=\frac{3}{\omega_D^3}\,\omega^{2}\propto\omega^{2},
\end{equation}
where $A_D=3/\omega_D^3$ is the Debye level and $\omega_D$ is the Debye frequency, given by
\begin{equation}
\omega_D=\left(\frac{c_L^{-3}+2c_T^{-3}}{3}\right)^{-1/3} q_D
=\left(\frac{18\pi^2\hat{\rho}}{c_L^{-3}+2c_T^{-3}}\right)^{1/3},
\end{equation}
with the Debye wavenumber $q_D=(6\pi^2\hat{\rho})^{1/3}$.
The resulting values of $c_L$, $c_T$, $q_D$, $A_D$, and $\omega_D$ for each model glass are summarized in Table~\ref{table2}.

%%%%%%%%%%%%%%%%%%%%%%%%%%%%%%%%%%%%%%%%%%%%%%%%%%%%%%%%%%%%%%%%%%%%%%%%%%%%%%%%%%%%%%%%%%%%%%%%%
\subsection{Elastic moduli}
We compute the elastic moduli---the bulk modulus $K$ and the shear modulus $G$---within the harmonic (athermal) linear-response framework~\cite{Lemaitre_2006,Mizuno3_2016,MizunoIkeda2022}.
These moduli determine the longitudinal and transverse sound speeds, and thereby the Debye quantities used throughout this work, including the Debye frequency $\omega_D$ and the Debye level $A_D$.

Table~\ref{table2} summarizes $K$, $G$, and the derived Debye parameters for the three model glasses studied here (silica, LJ, and SS).
In amorphous solids, elastic deformation generally contains both affine and nonaffine components~\cite{Lemaitre_2006,Zaccone_2011,Mizuno_2013,Mizuno3_2016}, such that an elastic modulus can be decomposed as $M=M_A-M_N$, where $M_A$ and $M_N$ denote the affine and nonaffine contributions, respectively.
We also report in Table~\ref{table2} the Poisson ratio $\nu=(3K-2G)/(6K+2G)$, which is approximately $\nu\simeq 0.2$ for silica glass, compared with $\nu\simeq 0.4$ for LJ and SS glasses.
Following the empirical correlation discussed in Refs.~\cite{Greaves_2011,Duval_2013}, values near $\nu\approx 0.2$ are characteristic of ``strong'' glasses, whereas $\nu\approx 0.4$ is typical of ``fragile'' glasses.

%%%%%%%%%%%%%%%%%%%%%%%%%%%%%%%%%%%%%%%%%%%%%%%%%%%%%%%
\begin{figure*}[t]
\centering
\includegraphics[width=1.0\textwidth]{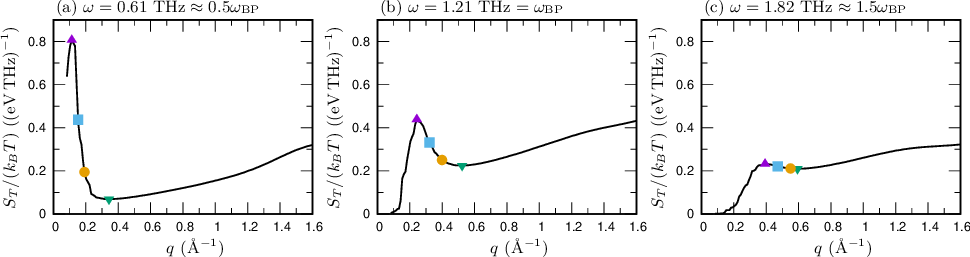}
\caption{\label{figure0}
{Crossover wavenumber $q_T(\omega)$ between phononic and non-phononic excitations in silica glass.}
The transverse dynamical structure factor $S_T(q,\omega)/(k_B T)$ is plotted as a function of wavenumber $q$ for three representative frequencies:
(a) $\omega=0.61~\mathrm{THz}\approx0.5\,\omega_{\mathrm{BP}}$,
(b) $\omega=1.21~\mathrm{THz}=\omega_{\mathrm{BP}}$, and
(c) $\omega=1.82~\mathrm{THz}\approx1.5\,\omega_{\mathrm{BP}}$.
Orange circles indicate $q_T(\omega)$ for each frequency.
Purple upward triangles and green downward triangles mark the wavenumbers $q_{\max}$ and $q_{\min}$ at which $S_T(q,\omega)$ takes its local maximum $S_{\max}$ and local minimum $S_{\min}$, respectively.
The cyan squares indicate the wavenumber $q_{\max}+q_{\mathrm{half}}$, where $q_{\mathrm{half}}$ is defined as the half-width measured from the phonon-branch peak such that $S_T(q,\omega)$ has decreased from $S_{\max}$ by half the peak-to-trough amplitude, $(S_{\max}-S_{\min})/2$.
The crossover wavenumber is then defined as $q_T(\omega)=q_{\max}+2\,q_{\mathrm{half}}$, and is shown by the orange circles.
}
\end{figure*}
%%%%%%%%%%%%%%%%%%%%%%%%%%%%%%%%%%%%%%%%%%%%%%%%%%%%%%%

%%%%%%%%%%%%%%%%%%%%%%%%%%%%%%%%%%%%%%%%%%%%%%%%%%%%%%%%%%%%%%%%%%%%%%%%%%%%%%%%%%%%%%%%%%%%%%%%%
\subsection{Dynamical structure factor}
Using the mode-resolved information $\{\omega_k,\vec{e}_k\}$, we compute the dynamical structure factors $S_\alpha(q,\omega)$, with $\alpha\in\{T,L\}$ labeling transverse ($\alpha=T$) and longitudinal ($\alpha=L$) components, following Refs.~\cite{Grest_1982,MizunoIkeda2022}:
\begin{equation}\label{eq.sqomega}
S_\alpha(q,\omega)=\frac{k_B T}{2N}\frac{q^2}{\omega^2}\sum_{k=1}^{3N} F_{k,\alpha}(q)\,\delta\!\left(\omega-\omega_k\right),
\end{equation}
where
\begin{equation}\label{eq.sqomega2}
\begin{aligned}
F_{k,T}(q)&=\left|\sum_{i=1}^{N}\left(\frac{\vec{e}_{k,i}}{\sqrt{m_i}}\times\hat{\vec{q}}\right)\exp\!\big(\mathrm{i}\vec{q}\cdot\vec{r}_i\big)\right|^{2},\\
F_{k,L}(q)&=\left|\sum_{i=1}^{N}\left(\frac{\vec{e}_{k,i}}{\sqrt{m_i}}\cdot\hat{\vec{q}}\right)\exp\!\big(\mathrm{i}\vec{q}\cdot\vec{r}_i\big)\right|^{2}.
\end{aligned}
\end{equation}
Here $k_B$ is the Boltzmann constant, $\vec{q}$ is the wavevector, $q=|\vec{q}|$, and $\hat{\vec{q}}=\vec{q}/q$.
Scattering experiments predominantly access the longitudinal component $S_L(q,\omega)$~\cite{Suck_1992}.

%%%%%%%%%%%%%%%%%%%%%%%%%%%%%%%%%%%%%%%%%%%%%%%%%%%%%%%%%%%%%%%%%%%%%%%%%%%%%%%%%%%%%%%%%%%%%%%%%
\subsubsection{Incoherent-scattering route to the vDOS}~\label{method.first}
As described in the Introduction, a practical route to estimate the vDOS from scattering observables is to use the large-wavenumber regime $q > q_D$, where the measured spectrum becomes effectively single-particle-like and can be related to $g(\omega)$ within an incoherent-scattering approximation.
Focusing on the longitudinal dynamical structure factor, which is the component most directly accessed in scattering experiments~\cite{Suck_1992}, $S_L(q,\omega)$ can be approximated within an incoherent-scattering (self) approximation as
\begin{equation} \label{eq_vdosexp_inc2}
\begin{aligned}
S_L(q,\omega)
&\approx
\frac{k_B T}{2N}\frac{q^2}{\omega^2}\sum_{k=1}^{3N}\sum_{i=1}^{N}
\left(\frac{\vec{e}_{k,i}}{\sqrt{m_i}}\cdot\hat{\vec{q}}\right)^{2}
\delta\!\left(\omega-\omega_k\right), \\
&\approx
\frac{k_B T}{2N}\frac{q^2}{\omega^2}\sum_{k=1}^{3N}
\sum_{i=1}^{N}\frac{\left|\vec{e}_{k,i}\right|^{2}}{3 m_i}
\delta\!\left(\omega-\omega_k\right), \\
&\approx
\frac{k_B T}{2}\frac{q^2}{\omega^2}\,M(\omega)^{-1}\,g(\omega),
\end{aligned}
\end{equation}
where we used the angular average over the direction of $\hat{\vec{q}}$, $\langle (\vec{a}\cdot\hat{\vec{q}})^2\rangle_{\hat{\vec{q}}}=|\vec{a}|^2/3$, in the second line.
The effective mass is defined through $M^{-1}=\sum_{i=1}^{N}|\vec{e}_{k,i}|^{2}/m_i$ evaluated at $\omega=\omega_k$; thus $M(\omega)$ is mode dependent and therefore frequency dependent.
This $M(\omega)$ accounts for the mass weighting of eigenmodes in multi-component systems.
For one-component systems, $M(\omega)$ reduces to the particle mass, $M(\omega)=m$.
We then obtain
\begin{equation} \label{eq_vdosexp_inc}
g_{\mathrm{inc}}(\omega)=2M(\omega)\omega^2\,\frac{S_L(q,\omega)}{q^2 k_B T}.
\end{equation}
An analogous derivation for the transverse channel $S_T(q,\omega)$ yields
\begin{equation}
g_{\mathrm{inc}}(\omega)=M(\omega)\omega^2\,\frac{S_T(q,\omega)}{q^2 k_B T}.
\end{equation}

In practice, to improve the accuracy of the incoherent-scattering approximation, we compute $g_{\mathrm{inc}}(\omega)$ by averaging the right-hand side of Eq.~(\ref{eq_vdosexp_inc}) over a high-$q$ window $q_1\le q\le q_2$~\cite{Price_1987,Pasquarello_1998,Fultz_2010,Squires_2012}, which corresponds to Eq.~(\ref{eq_vdosexp}).
This high-$q$ strategy exploits scattering dominated by essentially single-particle vibrations and thus provides a reconstruction of the vDOS in close analogy with the standard estimate based on the Fourier transform of the velocity autocorrelation function~\cite{Grest_1981,Ikeda_2013}.

%%%%%%%%%%%%%%%%%%%%%%%%%%%%%%%%%%%%%%%%%%%%%%%%%%%%%%%%%%%%%%%%%%%%%%%%%%%%%%%%%%%%%%%%%%%%%%%%%
\subsubsection{Wavenumber-resolved route to the vDOS}~\label{method.second}
As a second route, we access the vDOS by integrating the low-$q$ regime $q<q_D$, as outlined below.
Integrating Eqs.~(\ref{eq.sqomega})--(\ref{eq.sqomega2}) over $q\in[0,q_D]$ gives
\begin{equation}\label{eq.vdossqo1}
\begin{aligned}
&\int_{0}^{q_D}\!\!\left\{\frac{S_T(q,\omega)}{k_B T}+\frac{S_L(q,\omega)}{k_B T}\right\}\,dq \\
&= \frac{1}{2N\omega^{2}}\sum_{k=1}^{3N}\delta\!\left(\omega-\omega_k\right)
\left(\sum_{i,j=1}^{N}\frac{\vec{e}_{k,i}\cdot\vec{e}_{k,j}}{\sqrt{m_i m_j}}
\int_{0}^{q_D}\!e^{\mathrm{i}\vec{q}\cdot(\vec{r}_i-\vec{r}_j)}q^{2}\,dq\right).
\end{aligned}
\end{equation}
We now apply an isotropic-medium approximation.
In an isotropic elastic continuum, the eigenmodes are plane-wave elastic excitations,
\begin{equation}
\vec{e}_{k,i} \propto \vec{s}_\alpha\,e^{\mathrm{i}\vec{q}\cdot\vec{r}_i},
\end{equation}
where $\alpha\in\{T,L\}$ labels transverse ($\alpha=T$) and longitudinal ($\alpha=L$) polarizations, and $\vec{s}_\alpha$ is the corresponding polarization unit vector~\cite{MizunoIkeda2022}.
Plane waves with wavevectors restricted to $0\le|\vec{q}|\le q_D$ form an orthogonal and complete basis within the Debye cutoff.
Accordingly, one may use the completeness relation
\begin{equation}\label{eq:pw_completeness}
\sum_{\substack{\vec{q}:\,0\le|\vec{q}|\le q_D\\ \alpha\in\{T,L\}}}
\left(\vec{s}_\alpha e^{\mathrm{i}\vec{q}\cdot\vec{r}_i}\right)\cdot
\left(\vec{s}_\alpha^{\ast} e^{-\mathrm{i}\vec{q}\cdot\vec{r}_j}\right)
=
\sum_{\vec{q}:\,0\le|\vec{q}|\le q_D}
e^{\mathrm{i}\vec{q}\cdot (\vec{r}_i - \vec{r}_j)}
\;\propto\;
\delta_{i,j},
\end{equation}
where ${}^\ast$ denotes complex conjugation and we used $\sum_{\alpha\in\{T,L\}}\vec{s}_\alpha\cdot\vec{s}_\alpha^\ast=1$~\cite{MizunoIkeda2022}.
Under this isotropic-medium approximation [Eq.~(\ref{eq:pw_completeness})], the $q$-integral in Eq.~(\ref{eq.vdossqo1}) simplifies to
\begin{equation}\label{eq.isoem}
\begin{aligned}
&\int_{0}^{q_D}\!e^{\mathrm{i}\vec{q}\cdot(\vec{r}_i-\vec{r}_j)}q^{2}\,dq
=\frac{1}{4\pi}\int_{0\le|\vec{q}|\le q_D}\!e^{\mathrm{i}\vec{q}\cdot(\vec{r}_i-\vec{r}_j)}\,d^{3}\vec{q}\\
&=\frac{1}{4\pi}\delta_{i,j}\!\left(\int_{0\le|\vec{q}|\le q_D}\!d^{3}\vec{q}\right)
=\frac{q_D^{3}}{3}\,\delta_{i,j}.
\end{aligned}
\end{equation}
Substituting Eq.~(\ref{eq.isoem}) into Eq.~(\ref{eq.vdossqo1}) yields
\begin{equation} \label{eq.vdossqo2}
\begin{aligned}
&\int_{0}^{q_D}\!\!\left\{\frac{S_T(q,\omega)}{k_B T}+\frac{S_L(q,\omega)}{k_B T}\right\}\,dq \\
&\approx \frac{q_D^{3}}{2\omega^{2}}\frac{1}{3N}\sum_{k=1}^{3N}\delta\!\left(\omega-\omega_k\right)
\left(\sum_{i=1}^{N}\frac{|\vec{e}_{k,i}|^{2}}{m_i}\right)
= \frac{q_D^{3}}{2M(\omega)}\,\frac{g(\omega)}{\omega^{2}},
\end{aligned}
\end{equation}
where $M(\omega)$ is the effective mass defined in Eq.~(\ref{eq_vdosexp_inc2}).
We then obtain
\begin{equation} \label{eq_vdostl_method}
\begin{aligned}
\frac{g_{T+L}(\omega)}{\omega^2}
&=
\frac{2 M(\omega)}{q_D^3}
\int_0^{q_D}
\left[
\frac{S_T(q,\omega)}{k_B T}
+
\frac{S_L(q,\omega)}{k_B T}
\right]dq,
\end{aligned}
\end{equation}
which corresponds to Eq.~(\ref{eq_vdostl}).

In theoretical treatments~\cite{Schirmacher_2006,Marruzzo_2013,Schirmacher_2015,Mizuno_2018,Wyart_2010,Degiuli_2014}, Eq.~(\ref{eq_vdostl_method}) is used to evaluate the vDOS from the dynamical structure factor (or, equivalently, from the corresponding Green's functions).
A key advantage of this low-$q$ route is that it retains a decomposition into distinct wavenumber sectors, thereby providing a wavenumber-resolved link between the measured $S_\alpha(q,\omega)$ and the spatial character of vibrational excitations and their contributions to the BP.

%%%%%%%%%%%%%%%%%%%%%%%%%%%%%%%%%%%%%%%%%%%%%%%%%%%%%%%%%%%%%%%%%%%%%%%%%%%%%%%%%%%%%%%%%%%%%%%%%
\subsubsection{Crossover wavenumber $q_T(\omega)$ between phononic and non-phononic excitations}~\label{sec.qtcal}
As shown in Fig.~\ref{figure3}, the transverse dynamical structure factor $S_T(q,\omega)$ exhibits a phonon branch with an approximately linear dispersion at low $q$, and a non-phononic, nearly dispersionless band at higher $q$.
In Sec.~\ref{sec.phnph}, Eq.~(\ref{eq_vdos_decomp_T}) introduces a crossover wavenumber $q_T(\omega)$ to separate, for each frequency $\omega$, the vDOS into phononic and non-phononic contributions.
Here we describe how $q_T(\omega)$ is determined in practice.

We determine $q_T(\omega)$ from $S_T(q,\omega)$ viewed as a function of $q$ at fixed $\omega$.
Figure~\ref{figure0} plots $S_T(q,\omega)/(k_B T)$ versus $q$ for three representative frequencies, including $\omega_{\mathrm{BP}}$ as well as frequencies below and above $\omega_{\mathrm{BP}}$.
Orange circles indicate the resulting $q_T(\omega)$ for each frequency, obtained as follows.
(i) We first identify the phonon-branch peak, where $S_T(q,\omega)$ attains a local maximum $S_{\max}$ at wavenumber $q_{\max}$ (purple upward triangles).
(ii) We then locate the wavenumber $q_{\min}$ at which $S_T(q,\omega)$ reaches a local minimum $S_{\min}$ after the phonon-branch feature has decayed and the spectrum enters the dispersionless-band regime (green downward triangles).
(iii) Next, we determine $q_{\mathrm{half}}$ as a half-width measured from $q_{\max}$ such that $S_T(q,\omega)$ at $q=q_{\max}+q_{\mathrm{half}}$ has decreased from $S_{\max}$ by half the peak-to-trough amplitude, \textit{i.e.}, by $(S_{\max}-S_{\min})/2$ (cyan squares).
(iv) Finally, we define the crossover wavenumber as
$q_T(\omega)=q_{\max}+2\,q_{\mathrm{half}}$,
shown by the orange circles in Fig.~\ref{figure0}.

The resulting $q_T(\omega)$ is shown as the green curve in Fig.~\ref{figure3}.
Although this construction is not unique, Fig.~\ref{figure3} shows that the resulting $q_T(\omega)$ provides a reasonable and physically motivated boundary between the phonon-branch region and the non-phononic excitations.
We therefore adopt this definition throughout the present work.

%%%%%%%%%%%%%%%%%%%%%%%%%%%%%%%%%%%%%%%%%%%%%%%%%%%%%%%
\begin{figure*}[t]
\centering
\includegraphics[width=1.0\textwidth]{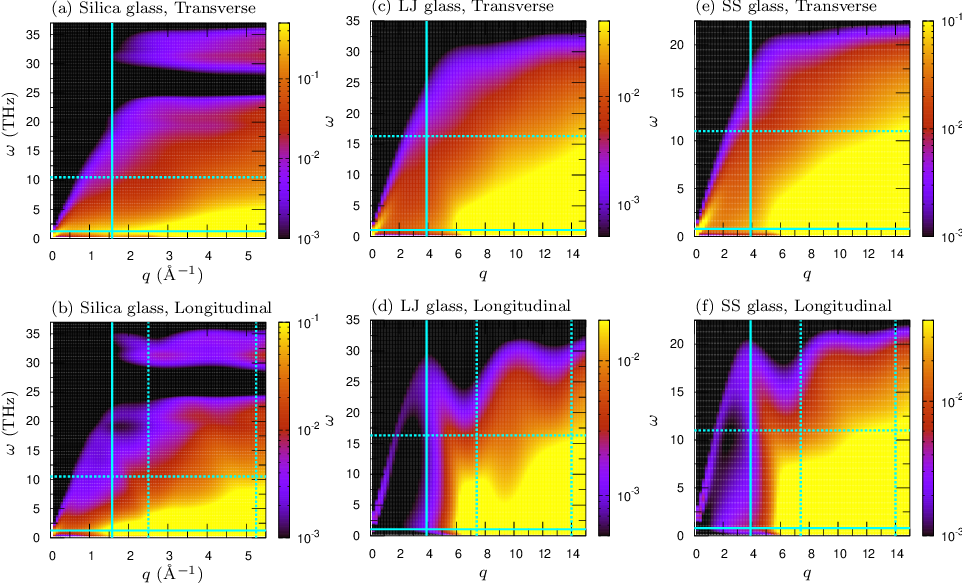}
\caption{\label{figure1}
{Dynamical structure factors.}
(a,b) Silica glass, (c,d) LJ glass, and (e,f) SS glass.
The transverse component $S_T(q,\omega)/(k_B T)$ is shown as a function of $q$ and $\omega$ in upper panels (a,c,e), and the longitudinal component $S_L(q,\omega)/(k_B T)$ in lower panels (b,d,f).
For silica glass, values are reported in units of $(\mathrm{eV}\,\mathrm{THz})^{-1}$.
The horizontal solid and dashed lines indicate the BP frequency $\omega_\text{BP}$ and the Debye frequency $\omega_D$, respectively.
The vertical solid line marks the Debye wavenumber $q_D$, while the vertical dashed lines in (b,d,f) mark $q_1$ and $q_2$ used to compute $g_\text{inc}(\omega)$ in Eq.~(\ref{eq_vdosexp}).
}
\end{figure*}
%%%%%%%%%%%%%%%%%%%%%%%%%%%%%%%%%%%%%%%%%%%%%%%%%%%%%%%

%%%%%%%%%%%%%%%%%%%%%%%%%%%%%%%%%%%%%%%%%%%%%%%%%%%%%%%
\begin{figure*}[t]
\centering
\includegraphics[width=1.0\textwidth]{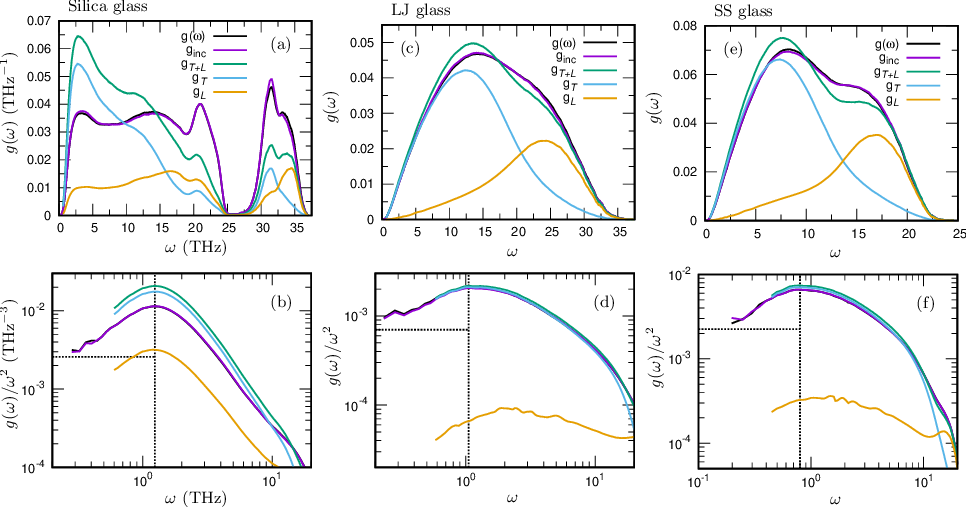}
\caption{\label{figure2}
{Vibrational density of states.}
(a,b) Silica glass, (c,d) LJ glass, and (e,f) SS glass.
$g(\omega)$ and $g(\omega)/\omega^2$ (black) are shown as functions of frequency $\omega$, together with $g_\text{inc}(\omega)$ (purple), calculated from $S_L(q,\omega)$ over the high-wavenumber range $q_1 \le q \le q_2$ (with $q_1,\ q_2 > q_D$) using Eq.~(\ref{eq_vdosexp}).
Also shown are the transverse and longitudinal components $g_T(\omega)$ (cyan) and $g_L(\omega)$ (orange), obtained by integrating $S_T(q,\omega)$ and $S_L(q,\omega)$ over $q \le q_D$ as in Eq.~(\ref{eq_vdostl}), and their sum $g_{T+L}(\omega)=g_T(\omega)+g_L(\omega)$ (green).
The vertical and horizontal dotted lines in (b,d,f) indicate the BP frequency $\omega_\text{BP}$ and the Debye level $A_D$, respectively.
}
\end{figure*}
%%%%%%%%%%%%%%%%%%%%%%%%%%%%%%%%%%%%%%%%%%%%%%%%%%%%%%%

%%%%%%%%%%%%%%%%%%%%%%%%%%%%%%%%%%%%%%%%%%%%%%%%%%%%%%%%%%%%%%%%%%%%%%%%%%%%%%%%%%%%%%%%%%%%%%%%%
\section{Results}
%
%%%%%%%%%%%%%%%%%%%%%%%%%%%%%%%%%%%%%%%%%%%%%%%%%%%%%%%%%%%%%%%%%%%%%%%%%%%%%%%%%%%%%%%%%%%%%%%%%
\subsection{Dynamical structure factor}
Figure~\ref{figure1} shows the simulated dynamical structure factors for the three glasses.
The upper panels (a,c,e) display the transverse component $S_T(q,\omega)$, whereas the lower panels (b,d,f) show the longitudinal component $S_L(q,\omega)$.
Focusing first on $S_T(q,\omega)$, silica, LJ, and SS glasses all exhibit qualitatively similar dependences on $q$ and $\omega$ in the low-frequency regime up to the Debye frequency $\omega_D$.
In particular, in the low-$\omega$ region around the BP frequency, $S_T(q,\omega)$ displays a linearly dispersing acoustic branch together with an additional, nearly $q$-independent (dispersionless) excitation band, which we discuss in detail below (see Fig.~\ref{figure3}).

The longitudinal channel, however, highlights a marked difference between the network glass (silica) and the packing-dominated glasses (LJ and SS).
For silica, $S_L(q,\omega)$ closely mirrors $S_T(q,\omega)$: besides the longitudinal acoustic branch, a dispersionless band is also visible in the low-$\omega$ region near the BP frequency.
For LJ and SS glasses, by contrast, $S_L(q,\omega)$ differs qualitatively from $S_T(q,\omega)$; the dispersionless band prominent in $S_T(q,\omega)$ is not apparent in $S_L(q,\omega)$, while the longitudinal acoustic branch persists to substantially higher frequencies, up to $\omega\simeq 20$ for LJ and $\omega\simeq 13$ for SS.
These observations suggest that transverse and longitudinal excitations are more strongly coupled in silica glass, whereas they are effectively decoupled in LJ and SS glasses.

This transverse--longitudinal coupling/decoupling trend is consistent with the nonaffine character of the elastic response, as summarized in Table~\ref{table2}.
In packing-dominated LJ and SS glasses, the shear modulus $G$ exhibits a sizable nonaffine reduction (about $60\%$ of the affine contribution), whereas the bulk modulus $K$ remains close to its affine value, with a nonaffine correction of order $1\%$; thus transverse (shear) and longitudinal (compressional) deformations are affected very differently by nonaffinity.
In contrast, in the covalent network glass (silica) both the shear and bulk moduli display strong nonaffine contributions, with nonaffine relaxations substantially reducing the moduli from their affine values and the nonaffine component reaching roughly $75\%$ of the affine one for both moduli.
This pronounced nonaffinity is brought about by the absence of local centrosymmetry in tetrahedral network structures~\cite{Milkus_2016,Krausser_2017} and has also been reported for amorphous silicon~\cite{Minamitani_2025} and noncentrosymmetric crystals such as $\alpha$-quartz~\cite{Cui_2019}.
As a consequence, $K$ and $G$ become closer in magnitude in silica glass than in packing-type glasses.

Taken together, these results indicate that the degree of transverse--longitudinal coupling in $S_\alpha(q,\omega)$ correlates with the extent of nonaffinity in the corresponding elastic moduli.
In network glasses, strong nonaffinity in both shear and bulk responses is accompanied by similar transverse and longitudinal spectral features.
In packing-dominated glasses, the near-affine bulk response is accompanied by a longitudinal spectrum in which the acoustic branch persists to substantially higher frequencies, with little trace of the transverse dispersionless band.

%%%%%%%%%%%%%%%%%%%%%%%%%%%%%%%%%%%%%%%%%%%%%%%%%%%%%%%%%%%%%%%%%%%%%%%%%%%%%%%%%%%%%%%%%%%%%%%%%
\subsection{Incoherent-scattering route to the vDOS}
As a first route, we access the vDOS from the dynamical structure factor in the large-wavenumber regime.
Concretely, we adopt the incoherent-scattering approximation and use the longitudinal component $S_L(q,\omega)$ in a high-$q$ window $q_1 \le q \le q_2$ to define~\cite{Price_1987,Pasquarello_1998,Fultz_2010,Squires_2012} [see Eq.~(\ref{eq_vdosexp_inc})]
\begin{equation}\label{eq_vdosexp}
g_{\mathrm{inc}}(\omega)
=
2M(\omega)\,\omega^2
\left\{
\frac{1}{q_2-q_1}\int_{q_1}^{q_2}\frac{S_L(q,\omega)}{q^2 k_B T}\,dq
\right\}.
\end{equation}
For silica glass, we choose $q_1=2.5$~\AA$^{-1}$ and $q_2=5.25$~\AA$^{-1}$, both larger than the Debye wavenumber $q_D=1.58$~\AA$^{-1}$, as indicated in Fig.~\ref{figure1}(b), so that the analysis is performed well within the large-$q$ regime.
For LJ and SS glasses, we likewise select $q_1$ and $q_2$ in the regime $q>q_D$; the specific choices are given together with the corresponding datasets in Fig.~\ref{figure1}(d,f).

Figure~\ref{figure2} compares $g_{\mathrm{inc}}(\omega)$ obtained from Eq.~(\ref{eq_vdosexp}) (purple curve) with the exact vDOS $g(\omega)$ computed from the full set of eigenfrequencies via Eq.~(\ref{eq.vdos}) (black curve).
In the lower panels [Fig.~\ref{figure2}(b,d,f)], we emphasize the low-frequency regime by plotting the reduced vDOS $g(\omega)/\omega^2$, which highlights the BP anomaly.
We find that $g_{\mathrm{inc}}(\omega)$ faithfully reproduces $g(\omega)$ over the full frequency range, including the BP feature and its characteristic frequency $\omega_\text{BP}$.
These results validate the large-$q$/incoherent-scattering route as a reliable procedure for extracting $g(\omega)$ from $S_L(q,\omega)$.

A further practical advantage is that scattering experiments predominantly probe the longitudinal channel $S_L(q,\omega)$ and often yield especially robust statistics in the large-$q$ regime.
Accordingly, Eq.~(\ref{eq_vdosexp}) provides a convenient experimental pathway to estimate $g(\omega)$ directly from high-$q$ measurements of $S_L(q,\omega)$.

%%%%%%%%%%%%%%%%%%%%%%%%%%%%%%%%%%%%%%%%%%%%%%%%%%%%%%%%%%%%%%%%%%%%%%%%%%%%%%%%%%%%%%%%%%%%%%%%%
\subsection{Wavenumber-resolved route to the vDOS}
Next, we focus on the second, wavenumber-resolved route to the vDOS.
In this approach, $g(\omega)$ is estimated by integrating the dynamical structure factor over wavenumbers from $0$ to $q_D$.
Specifically, we use [see Eq.~(\ref{eq_vdostl_method})]
\begin{equation}\label{eq_vdostl}
\begin{aligned}
\frac{g_{T+L}(\omega)}{\omega^2}
&=
\frac{2 M(\omega)}{q_D^3}
\int_0^{q_D}
\left[
\frac{S_T(q,\omega)}{k_B T}
+
\frac{S_L(q,\omega)}{k_B T}
\right]dq, \\
&\equiv
\frac{g_T(\omega)}{\omega^2}
+
\frac{g_L(\omega)}{\omega^2}.
\end{aligned}
\end{equation}
This expression naturally decomposes the vDOS into transverse and longitudinal contributions, $g_T(\omega)$ and $g_L(\omega)$, defined by integrating $S_T(q,\omega)$ and $S_L(q,\omega)$ over $0< q<q_D$, respectively.
In theoretical treatments~\cite{Schirmacher_2006,Marruzzo_2013,Schirmacher_2015,Mizuno_2018,Wyart_2010,Degiuli_2014}, Eq.~(\ref{eq_vdostl}) is commonly used to evaluate the vDOS from the dynamical structure factor (or, equivalently, from the corresponding Green's functions).

Figure~\ref{figure2} shows the transverse $g_T(\omega)$ and longitudinal $g_L(\omega)$, together with their sum $g_{T+L}(\omega)\equiv g_T(\omega)+g_L(\omega)$
\footnote{
In the second, wavenumber-resolved approach based on Eq.~(\ref{eq_vdostl}), accurately estimating the low-frequency vDOS requires reliable data for the phonon-branch contribution in the low-$q$ part of the dynamical structure factor.
If the low-$q$ phonon branch is not computed with sufficient accuracy, the $q$-integration in Eq.~(\ref{eq_vdostl}) becomes unreliable at small $\omega$, making it difficult to obtain a quantitative estimate of $g_{T+L}(\omega)$ near the low-frequency edge.
For this reason, in Fig.~\ref{figure2} we do not display the data at the lowest frequencies, where the vDOS estimate from the second route is not sufficiently controlled.
}.
For silica glass [panels (a,b)], $g_{T+L}(\omega)$ captures the overall $\omega$ dependence of the exact vDOS $g(\omega)$, although noticeable quantitative deviations remain.
For the packing-dominated glasses, Eq.~(\ref{eq_vdostl}) performs better: as shown in panels (c--f), $g_{T+L}(\omega)$ quantitatively reproduces $g(\omega)$ for the LJ and SS glasses.
The larger deviations in silica glass are plausibly attributed to its tetrahedral network topology, in contrast to the more randomly and tightly packed structures of the LJ and SS glasses, which may lead to departures from the isotropic-medium approximation [Eq.~(\ref{eq:pw_completeness})] and hence from Eq.~(\ref{eq_vdostl}).
Nevertheless, even for silica glass the estimate $g_{T+L}(\omega)$ from Eq.~(\ref{eq_vdostl}) captures the overall frequency dependence of the exact $g(\omega)$; in particular, in the BP regime it reproduces the characteristic $\omega$ dependence, including the BP frequency $\omega_\text{BP}$.
This indicates that Eq.~(\ref{eq_vdostl}) provides a useful basis for vDOS analysis within the present wavenumber-resolved route.

A further key observation in Fig.~\ref{figure2}(b,d,f) is that, in the low-$\omega$ region around $\omega_{\mathrm{BP}}$, the transverse contribution $g_T(\omega)$ substantially exceeds the longitudinal contribution $g_L(\omega)$ in all glasses studied here (compare the cyan and orange curves).
This indicates that the BP is predominantly associated with transverse vibrational states.
Such transverse dominance has been reported previously for packing-type glasses~\cite{Monaco2_2009,Mizuno2_2013,Mizuno_2014,Mizuno_2018} and also for network-type systems such as amorphous silicon~\cite{Beltukov_2016,Beltukov_2018,Minamitani_2022}.

Importantly, however, silica glass differs from the packing-dominated LJ and SS glasses in the relative weight of the longitudinal sector.
Notably, in silica glass, $g_L(\omega)$ is larger and closer to $g_T(\omega)$ than in the LJ and SS glasses, tracking the $\omega$ dependence of the exact $g(\omega)$ more faithfully, including the location of $\omega_{\mathrm{BP}}$ [panel (b)].
By contrast, this behavior is not observed for the LJ and SS glasses, where $g_L(\omega)$ remains much smaller than $g_T(\omega)$ in the BP regime.
This distinction mirrors the dynamical-structure-factor phenomenology in Fig.~\ref{figure1}:
in silica glass, the transverse and longitudinal channels exhibit similar low-$q$/low-$\omega$ spectral features (indicative of stronger transverse--longitudinal coupling), whereas in the LJ and SS glasses, the longitudinal spectrum shows a pronounced acoustic branch that persists to substantially higher frequencies, with little trace of the transverse dispersionless band (indicative of effective decoupling).

%%%%%%%%%%%%%%%%%%%%%%%%%%%%%%%%%%%%%%%%%%%%%%%%%%%%%%%
\begin{figure*}[t]
\centering
\includegraphics[width=1.0\textwidth]{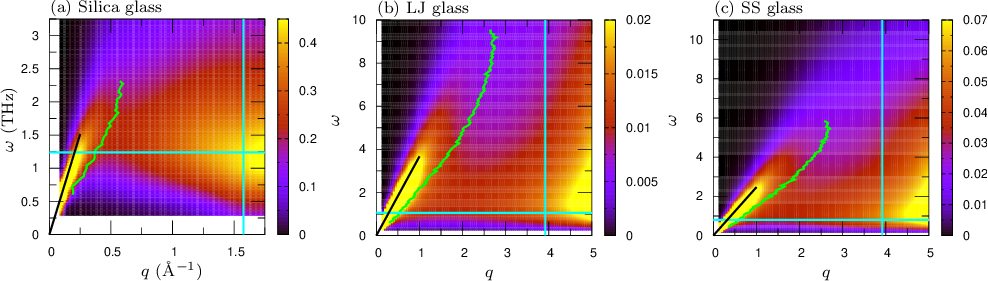}
\caption{\label{figure3}
{Transverse dynamical structure factor at low wavenumbers and low frequencies.}
(a) Silica glass, (b) LJ glass, and (c) SS glass.
$S_T(q,\omega)/(k_B T)$ is shown as a function of $q$ and $\omega$.
For silica glass, values are reported in units of $(\mathrm{eV}\,\mathrm{THz})^{-1}$.
The vertical line marks the Debye wavenumber $q_D$.
The horizontal line indicates the BP frequency $\omega_\text{BP}$.
The black curve shows the linear dispersion $\omega = c_T q$, where $c_T$ is the transverse sound speed, corresponding to phonon excitations.
The green curve indicates $q_T(\omega)$, used to separate $g_T(\omega)$ into the phononic component $g_{\mathrm{Ph}}(\omega)$ and the non-phononic component $g_{\mathrm{Nph}}(\omega)$ in Eq.~(\ref{eq_vdos_decomp_T}).
}
\end{figure*}
%%%%%%%%%%%%%%%%%%%%%%%%%%%%%%%%%%%%%%%%%%%%%%%%%%%%%%%

%%%%%%%%%%%%%%%%%%%%%%%%%%%%%%%%%%%%%%%%%%%%%%%%%%%%%%%
\begin{figure*}[t]
\centering
\includegraphics[width=1.0\textwidth]{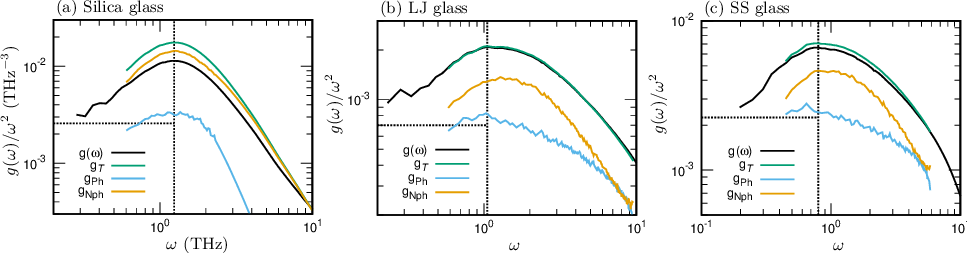}
\caption{\label{figure4}
{Phononic and non-phononic components of the vibrational density of states.}
(a) Silica glass, (b) LJ glass, and (c) SS glass.
The reduced vDOS $g(\omega)/\omega^2$ (black) is plotted as a function of the angular frequency $\omega$, together with the transverse contribution $g_T(\omega)/\omega^2$ (green), obtained by integrating $S_T(q,\omega)$ over $0\le q\le q_D$ according to Eq.~(\ref{eq_vdostl}).
We further decompose $g_T(\omega)/\omega^2$ into its phononic and non-phononic components, $g_{\mathrm{Ph}}(\omega)/\omega^2$ (cyan) and $g_{\mathrm{Nph}}(\omega)/\omega^2$ (orange), as defined in Eq.~(\ref{eq_vdos_decomp_T}).
The vertical and horizontal dotted lines mark the BP frequency $\omega_{\mathrm{BP}}$ and the Debye level $A_D$, respectively.
}
\end{figure*}
%%%%%%%%%%%%%%%%%%%%%%%%%%%%%%%%%%%%%%%%%%%%%%%%%%%%%%%

%%%%%%%%%%%%%%%%%%%%%%%%%%%%%%%%%%%%%%%%%%%%%%%%%%%%%%%%%%%%%%%%%%%%%%%%%%%%%%%%%%%%%%%%%%%%%%%%%
\subsection{Phononic and non-phononic components of the vDOS}\label{sec.phnph}
The key advantage of the second, wavenumber-resolved route to the vDOS is that it enables a quantitative decomposition of how each wavenumber sector of $S_T(q,\omega)$ and $S_L(q,\omega)$ contributes to the vDOS $g(\omega)$.
Exploiting this capability, we decompose the vDOS into contributions from the phonon branch and from the remaining non-phononic excitations, as described below.
As shown above, the transverse sector provides the dominant contribution in the BP regime, and we therefore focus on $S_T(q,\omega)$ and $g_T(\omega)$.

Figure~\ref{figure3} presents the transverse dynamical structure factor $S_T(q,\omega)$ in the low-$q$ and low-$\omega$ regime.
As reported in our recent work~\cite{Mizuno_2025}, silica, LJ, and SS glasses all exhibit, in addition to the phonon branch, a broad and nearly dispersionless excitation band in the BP-frequency regime that persists up to the Debye wavenumber $q_D$.
Moreover, Mahajan \textit{et al.}~\cite{Mahajan_2025} recently compiled existing experimental and simulation data and concluded that such dispersionless-band excitations occur across a wide variety of amorphous systems.
These observations suggest that the broad dispersionless band is a generic feature of amorphous solids and glasses.

We therefore separate the vDOS into the phonon-branch contribution and the contribution from the broad dispersionless (non-phononic) band.
For each frequency $\omega$, we introduce a crossover wavenumber $q=q_T(\omega)$ in $S_T(q,\omega)$ that separates the phonon-branch-dominated regime ($q<q_T(\omega)$) from the dispersionless-band-dominated regime ($q>q_T(\omega)$).
We then define the phononic and non-phononic contributions, $g_{\mathrm{Ph}}(\omega)$ and $g_{\mathrm{Nph}}(\omega)$, by
\begin{equation}\label{eq_vdos_decomp_T}
\begin{aligned}
\frac{g_{\mathrm{Ph}}(\omega)}{\omega^2}
&=
\frac{2 M(\omega)}{q_D^3}
\int_0^{q_{T}(\omega)}
\frac{S_T(q,\omega)}{k_B T}\,dq, \\
\frac{g_{\mathrm{Nph}}(\omega)}{\omega^2}
&=
\frac{2 M(\omega)}{q_D^3}
\int_{q_{T}(\omega)}^{q_D}
\frac{S_T(q,\omega)}{k_B T}\,dq .
\end{aligned}
\end{equation}
By construction, the transverse contribution satisfies $g_T(\omega)=g_{\mathrm{Ph}}(\omega)+g_{\mathrm{Nph}}(\omega)$.
The crossover wavenumber $q_T(\omega)$ is determined by analyzing $S_T(q,\omega)$ as a function of $q$ at fixed $\omega$.
Details of the procedure are given in Sec.~\ref{sec.qtcal}.
The resulting $q_T(\omega)$ is shown as the green curve in Fig.~\ref{figure3}.
As can be seen from the figure, this criterion provides a reasonable boundary separating the phonon branch from the broad dispersionless band.

Figure~\ref{figure4} shows the resulting decomposition into $g_{\mathrm{Ph}}(\omega)$ and $g_{\mathrm{Nph}}(\omega)$.
For all three glasses, the decomposition demonstrates quantitatively that, in the BP-frequency regime, the vDOS contains mixed contributions from phononic and non-phononic excitations.
The phononic component $g_{\mathrm{Ph}}(\omega)$ approximately follows the Debye form $A_D\omega^2$, as expected.
In addition, glasses exhibit a non-phononic component $g_{\mathrm{Nph}}(\omega)$, which constitutes the excess modes over the Debye prediction and thus underlies the BP anomaly.
These results provide direct, quantitative evidence that the broad band observed in the dynamical structure factor corresponds to non-phononic excitations and is the microscopic origin of the excess vDOS and the BP in both network-type and packing-type glasses.

%%%%%%%%%%%%%%%%%%%%%%%%%%%%%%%%%%%%%%%%%%%%%%%%%%%%%%%
\begin{figure}[t]
\centering
\includegraphics[width=0.425\textwidth]{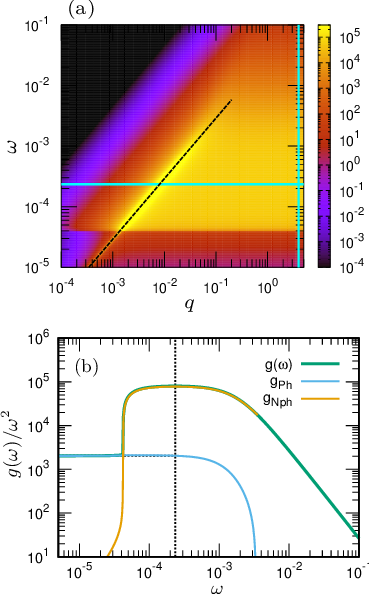}
\vspace*{-1.0mm}
\caption{\label{figure5}
Predictions of the EMT for random spring networks.
(a) The dynamical structure factor $S(q,\omega)/(k_B T)$ [Eq.~(\ref{emtsqo})] plotted as a function of wavenumber $q$ and frequency $\omega$.
(b) The reduced vDOS $g(\omega)/\omega^2$ [Eq.~(\ref{emtvdos})] plotted as a function of $\omega$, together with its phononic and non-phononic components, $g_{\mathrm{Ph}}(\omega)/\omega^2$ and $g_{\mathrm{Nph}}(\omega)/\omega^2$, calculated from Eq.~(\ref{emtphnph}).
In panel (a), the vertical line marks the Debye wavenumber $q_D$, and the horizontal line indicates the BP frequency $\omega_{\mathrm{BP}}$.
The black dashed curve shows the linear dispersion $\omega=cq$, where $c$ is the sound speed, corresponding to phonon excitations.
In panel (b), the vertical and horizontal dotted lines indicate $\omega_{\mathrm{BP}}$ and the Debye level $A_D$, respectively.
}
\end{figure}
%%%%%%%%%%%%%%%%%%%%%%%%%%%%%%%%%%%%%%%%%%%%%%%%%%%%%%%

%%%%%%%%%%%%%%%%%%%%%%%%%%%%%%%%%%%%%%%%%%%%%%%%%%%%%%%%%%%%%%%%%%%%%%%%%%%%%%%%%%%%%%%%%%%%%%%%%
\subsection{Effective medium theory}
Finally, we show that our simulation results are fully consistent with the predictions of the EMT for random spring networks~\cite{Wyart_2010,Degiuli_2014}, which provides a theoretical basis for interpreting the broad band and the BP in terms of isostaticity and marginal stability~\cite{Mizuno_2025}.
We perform an EMT analysis following Refs.~\cite{Wyart_2010,Degiuli_2014}, in which particles (nodes) are connected by linear springs to form a random spring network.
Two control parameters govern mechanical stability: the connectivity (coordination number) $z$ and the level of pre-stress $e>0$, which quantifies the internal forces carried by the springs.
Mechanical stability requires $z$ to exceed the isostatic threshold $z_c=2d$ in $d$ dimensions, \textit{i.e.}, a positive excess coordination $\delta z=z-z_c>0$.
In addition, if the pre-stress exceeds a critical value $e_c>0$, the network becomes unstable; thus stability requires $e<e_c$.
Within the EMT approximation, the random spring network is characterized by a complex, frequency-dependent effective spring constant $k_{\rm eff}(\omega)$ (see Refs.~\cite{Wyart_2010,Degiuli_2014} for derivations and the full formulation), from which both the vDOS $g(\omega)$ and the dynamical structure factor $S(q,\omega)$ can be computed, as detailed below.

Given $k_{\rm eff}(\omega)$, the vDOS and dynamical structure factor are obtained as
\begin{align}
g(\omega) &= \frac{2m\omega}{\pi}\,\mathrm{Im}\!\left[\frac{3}{q_D^3}\int_0^{q_D}\frac{q^2\,dq}{k_{\rm eff}(\omega)\,q^2 - m\omega^2}\right], \label{emtvdos} \\
S(q,\omega) &= \frac{k_B T}{\pi}\,\frac{q^2}{\omega}\,\mathrm{Im}\!\left[\frac{1}{k_{\rm eff}(\omega)\,q^2 - m\omega^2}\right], \label{emtsqo}
\end{align}
where $\mathrm{Im}$ denotes the imaginary part.
Because transverse and longitudinal polarizations are not distinguished in this EMT, one has $S(q,\omega)=S_L(q,\omega)=S_T(q,\omega)/2$.
Combining these expressions yields
\begin{equation}
\frac{g(\omega)}{\omega^{2}} = \frac{2m}{q_D^{3}} \int_{0}^{q_D} 3\,\frac{S(q,\omega)}{k_B T}\,dq,
\end{equation}
which corresponds to Eq.~(\ref{eq_vdostl}) upon identifying $3S\to S_T+S_L$ (sum over two transverse and one longitudinal polarizations) and, for identical particles in EMT, $m$ with the effective mass $M(\omega)$.
In what follows, we set $q_D=4$ (dimensionless EMT units) and $m=1$.

To capture the BP phenomenology observed in our simulations, we choose $\delta z=z-z_c=10^{-2}$ and $e=e_c(1-5\times10^{-4})$, placing the network close to both isostaticity ($\delta z\ll 1$) and marginal stability ($e\lesssim e_c$).
Figure~\ref{figure5}(a) shows the resulting dynamical structure factor $S(q,\omega)/(k_B T)$.
The spectrum exhibits phonon-like excitations with an approximately linear dispersion $\omega=cq$, where $c$ is the sound speed.
In addition, a broad, nearly wavenumber-independent (dispersionless) non-phononic band emerges around the BP regime, in direct analogy with the simulation results.

Figure~\ref{figure5}(b) presents the phononic and non-phononic contributions to the vDOS, $g_{\mathrm{Ph}}(\omega)$ and $g_{\mathrm{Nph}}(\omega)$.
Within EMT, these components are expressed in terms of $k_{\rm eff}(\omega)$ as
\begin{equation} \label{emtphnph}
\begin{aligned}
g_{\mathrm{Ph}}(\omega) &= \frac{2m\omega^2}{\pi}\,\frac{3\pi}{2q_D^3}\,\mathrm{Re}\!\left[\frac{1}{k_{\rm eff}(\omega)^{3/2}}\right], \\
g_{\mathrm{Nph}}(\omega) &= \frac{2m\omega}{\pi}\,\frac{3}{q_D^2}\,\mathrm{Im}\!\left[\frac{1}{k_{\rm eff}(\omega)}\right],
\end{aligned}
\end{equation}
where $\mathrm{Re}$ denotes the real part.
The phononic component $g_{\mathrm{Ph}}(\omega)$ approximately follows the Debye form $A_D\omega^2$, as expected, whereas the non-phononic component $g_{\mathrm{Nph}}(\omega)$ provides the excess modes over the Debye prediction and thus underlies the BP anomaly.

Taken together, the EMT predictions reproduce the same phenomenology as our simulations: a broad dispersionless band in $S(q,\omega)$ and a corresponding excess contribution $g_{\mathrm{Nph}}(\omega)$ that generates the BP.
This consistency supports a unified scenario in which the dispersionless band and the BP originate from the proximity of glasses to isostaticity and marginal stability.
Indeed, in our recent work~\cite{Mizuno_2025}, we demonstrated with more detailed simulation analyses that the BP in glasses can be rationalized within an isostaticity--marginal-stability framework.

%%%%%%%%%%%%%%%%%%%%%%%%%%%%%%%%%%%%%%%%%%%%%%%%%%%%%%%%%%%%%%%%%%%%%%%%%%%%%%%%%%%%%%%%%%%%%%%%%
\section{Conclusions}
In this study, we investigated BP anomalies in both a network-type glass (silica) and packing-type glasses (LJ and SS) by quantifying the vDOS from the dynamical structure factor using two complementary routes.
In the first route, we assume incoherent scattering and extract the vDOS from the high-wavenumber data of longitudinal $S_L(q,\omega)$~\cite{Price_1987,Pasquarello_1998,Fultz_2010,Squires_2012}.
We verified that this procedure reproduces the exact vDOS over the entire frequency range.
Because scattering experiments predominantly access the longitudinal channel $S_L(q,\omega)$~\cite{Suck_1992}, this route provides a practical experimental pathway to accurately determine the vDOS, including the BP.

In the second route, we quantify the vDOS by integrating the dynamical structure factors over low wavenumbers from $0$ up to the Debye wavenumber~\cite{Schirmacher_2006,Marruzzo_2013,Schirmacher_2015,Mizuno_2018,Wyart_2010,Degiuli_2014}.
Its key advantage is that it enables a quantitative decomposition of the vDOS into transverse and longitudinal sectors and, more importantly, into distinct wavenumber ranges.
Using this decomposition, we found that the BP is predominantly governed by transverse vibrational states in all glasses studied here (silica, LJ, and SS), irrespective of whether they are network-type or packing-type.
Furthermore, by exploiting the wavenumber resolution, we separated the vDOS into phononic and non-phononic parts.
We demonstrated quantitatively that, in the BP regime, the phononic contribution remains close to the Debye level, while an additional non-phononic contribution associated with the dispersionless excitation band builds up on top of it, thereby forming the BP anomaly.

While these conclusions hold universally across both network- and packing-type glasses, important differences also emerge between silica and the packing-dominated glasses.
In silica glass, the tetrahedral network breaks local centrosymmetry, which gives rise to strong nonaffine responses in both shear and bulk deformations~\cite{Milkus_2016,Krausser_2017,Minamitani_2025,Cui_2019}.
Consistent with this elastic behavior, the transverse and longitudinal dynamical structure factors are more strongly coupled in silica, and $S_L(q,\omega)$ exhibits a $q$--$\omega$ dependence qualitatively similar to that of $S_T(q,\omega)$.
As a result, insights obtained from the transverse channel can be more directly carried over to the longitudinal channel in silica glass.
From an experimental standpoint, this is a particularly important advantage, because scattering measurements typically provide access to the longitudinal component~\cite{Suck_1992}.

Finally, our simulation results are fully consistent with the effective-medium theory for random spring networks~\cite{Wyart_2010,Degiuli_2014}.
This agreement indicates that both the BP and the dispersionless excitation band can be understood within a unified framework based on isostaticity and marginal stability~\cite{Mizuno_2025}.
Building on these insights, a systematic re-examination of existing experimental and simulation datasets should further strengthen the present scenario and clarify its broader applicability~\cite{Mahajan_2025}.

%%%%%%%%%%%%%%%%%%%%%%%%%%%%%%%%%%%%%%%%%%%%%%%%%%%%%%%%%%%%%%%%%%%%%%%%%%%%%%%%%%%
\section*{Acknowledgments}
We thank Tatsuya Mori and Maiko Kofu for useful discussions.
This work was supported by JSPS KAKENHI Grant Nos.~23H04495 and 25H01519, JST FOREST Grant No. JPMJFR236Q, and JST ERATO Grant No.~JPMJER2401.

%%%%%%%%%%%%%%%%%%%%%%%%%%%%%%%%%%%%%%%%%%%%%%%%%%%%%%%%%%%%%%%%%%%%%%%%%%%%%%%%%%%
\section*{Author contributions statement}
H.M. designed the research, analyzed the data, and wrote the paper.
H.M and E.M. performed the research.

%%%%%%%%%%%%%%%%%%%%%%%%%%%%%%%%%%%%%%%%%%%%%%%%%%%%%%%%%%%%%%%%%%%%%%%%%%%%%%%%%%%
\section*{Author declarations}
The authors declare no conflicts of interest.

%%%%%%%%%%%%%%%%%%%%%%%%%%%%%%%%%%%%%%%%%%%%%%%%%%%%%%%%%%%%%%%%%%%%%%%%%%%%%%%%%%%%%%%%%%%%%%%%%%%%%%%%%%%%%%%%%%%%%%%%%%%
\bibliographystyle{apsrev4-2}
\bibliography{reference}

\end{document}